\documentstyle[12pt]{article}    

\def\be{\begin{equation}}
\def\ee{\end{equation}}
\def\beq{\begin{eqnarray}}
\def\eeq{\end{eqnarray}}

\def\bay{\begin{array}}
\def\eay{\end{array}}

\begin{document}


\title{Inflation driven by causal heat flux}

\author{
R. Maartens\dag\ddag, M. Govender\ddag\, and S.D. Maharaj\ddag
\\
\\
\dag {\footnotesize
School of Computer Science and Mathematics, 
Portsmouth University, Portsmouth
PO1 2EG, England} \\
\ddag {\footnotesize Department of Mathematics
and Applied Mathematics, University of Natal, Durban 4041, South 
Africa}
}

\date{October 1997}

\maketitle

\[ \]

\begin{abstract}

We find a simple inflationary solution in an inhomogeneous
spacetime with heat flux. The heat flux obeys a causal transport
equation, and counteracts the inflationary decrease of energy
density.
At late times, the heat flux tends to zero and the fluid approaches
the equation of state $p=-\rho$.

\end{abstract}
\[ \]




\section{Introduction}

Inflationary expansion arises when the effective pressure becomes
sufficiently negative that its `repulsive' contribution to
gravity becomes dominant. This scenario occurs naturally in
scalar-field models of the early universe. On the other hand, various
papers have considered the possibility that inflation could occur in
an `ordinary' fluid if there is sufficient bulk viscous stress to
drive the effective pressure negative (see, e.g., \cite{pbj,m,z} and
references given there). Clearly, as pointed out in \cite{mm}, the
bulk viscous stress must exceed the local equilibrium pressure, so
that such models are far from equilibrium.

Bulk viscous stress is a scalar dissipative effect, and thus
compatible with the spatial homogeneity and isotropy of
Friedmann-Lemaitre-Roberston-Walker (FLRW) spacetime. The vector
dissipative effect of heat flux is not compatible with
FLRW symmetry.
Here we present a `toy' model to show that, in a simple inhomogeneous
geometry, it is in principle possible for inflationary expansion to
arise in a fluid with heat flux. Essentially, the heat flux acts to
counter the decrease of energy density $\rho$ due to expansion, while
the pressure $p$ is steadily reduced (with asymptotic limit
$p\rightarrow -\rho$). Unlike the viscous inflationary models,
the fluid is not far from equilibrium at all spacetime events.

For consistency, we use the causal transport
equation of Israel and Stewart for heat flux, in
preference to the often used Fourier equation (generalized to
relativity), which is noncausal and has unstable equilibrium
states \cite{hl}.

\section{The basic equations}

Consider Modak's shear-free spherically symmetric model,
given in comoving coordinates by \cite{tp}
\be
ds^2 = -\left[1+M(t)r^2\right]^2dt^2 +
a^2(t)\left[dr^2 + r^2(d\theta^2 +
\sin^2{\theta}d\phi^2)\right]\,,  \label{1}
\ee
where the
fluid four-velocity is $u^\alpha = [1+Mr^2]^{-1}
\delta^\alpha{}_0$.
The fluid four-acceleration is
\be
a_\alpha\equiv u^\beta\nabla_\beta u_\alpha=\left({2Mr\over
1+Mr^2}\right)\delta_\alpha{}^1 \,,
\label{2}\ee
and the volume expansion rate is
\be
\Theta\equiv \nabla^\alpha u_\alpha=
{3H\over 1+Mr^2}\,,~~H\equiv{\dot{a}\over a}\,,
\label{3}\ee
where a dot denotes $\partial/\partial t$.
The heat flux vector is
\be
q^\alpha=\left({q\over a}\right)\delta^\alpha{}_1 ~~
\mbox{ where }~~q^2(t,r)=q^\alpha q_\alpha \,.
\label{12}\ee
Note that $q$ is a covariant scalar measure of the heat flux
magnitude.
The four-velocity $u^\alpha$
is comoving with the particle frame, in which the
total energy flux is the heat flux, since there is no particle flux
relative to this frame.

The Einstein field equations\footnote{We use
units with $8\pi G=1=c$}
are satisfied if
\beq
\rho &=& {\textstyle{1\over3}}\Theta^2 \,, \label{6a} \\
p &=& \displaystyle\frac{1}{(1 + Mr^2)^2}
\left[ {\textstyle{2\over3}}{\dot{M}}r^2 \Theta
-2{\ddot{a}\over a}+{4M\left(1+Mr^2\right)\over
a^2}\right]-{\textstyle{1\over3}}\rho \,,
\label{6b} \\
q &=& -\displaystyle\frac{4Mr\Theta}{3(1 + Mr^2)a} \,. \label{6c}
\eeq
Then the conservation equations
\beq
&& u^\alpha\nabla_\alpha
{\rho}+(\rho+p)\Theta+\nabla^\alpha q_\alpha+a^\alpha
q_\alpha =0\,, \label{4a}\\
&& (\rho+p)a_\alpha+h_\alpha{}^\beta\left(\nabla_\beta p+
u^\gamma\nabla_\gamma{q}_\beta\right)+{\textstyle{4\over3}}\Theta
q_\alpha=0 \,, \label{4b}
\eeq
are identically satisfied, where $h_{\alpha\beta}=g_{\alpha\beta}
+u_\alpha u_\beta$ projects orthogonal to $u^\alpha$, and
$g_{\alpha\beta}$ is the metric. In Modak's model, the heat flux
is radially inward if the fluid is expanding, and this is reflected
in the fact that the energy density at each instant of time is a
maximum at the centre of symmetry.

The functions $a(t)$ and $M(t)$
are determined after appropriate thermodynamic equations are imposed.
We now require that the heat flux is governed by
the causal transport equation of Israel and Stewart \cite{hl}
\be \label{7}
\tau h^{\alpha\beta} u^\gamma\nabla_\gamma
{q}_\beta+q^\alpha = -\kappa \left(
h^{\alpha\beta}\nabla_\beta T+Ta^\alpha\right)\,,
\ee
where $T$ is the local equilibrium
temperature, $\kappa$ ($\geq0$) is the thermal conductivity and
$\tau$ is the relaxational time-scale which gives rise to the
causal and stable behaviour of the theory. The noncausal
Fourier transport
equation has $\tau=0$ and reduces from an evolution equation to an
algebraic constraint on the heat flux. Intuitively, one can see that
in this case, the heat flux is instantaneously brought to zero when
the temperature gradient and acceleration are `switched off'.

For the line element (\ref{1}), equation (\ref{7}) becomes
\be \label{8}
\tau\dot{q} + \left(1+Mr^2\right)q =
-{\kappa\over a}
\left[\left(1+Mr^2\right)T\right]' \,,
\ee
where a prime denotes $\partial/\partial r$.

\section{A simple inflationary model}

To close the system of equations, one needs further
thermodynamic information about $\tau$, $\kappa$, $\rho$, $p$ and
$T$.
In \cite{tp}, this is done for a radiation-dominated model
where heat transport arises from radiative transfer. Here we follow a
different approach, since we are interested in demonstrating the
theoretical possibility of inflation driven by causal heat flux.
First we suppose that $M=M_0$ is a positive constant. Then
on each comoving sphere $r=$ constant of fluid particles,
inflationary
expansion is characterized by $\ddot{a}>0$. A particular case of
inflationary expansion is $H=H_0$ where $H_0$ is a positive constant,
and then $a=a_0\exp(H_0t)$.
Substituting into equations (\ref{6a})--(\ref{6c}), we find
\beq
\rho &=& {3H_0^2\over \left(1+M_0r^2\right)^2} \,, \label{5a}\\
p &=& \left[{4M_0\over a_0^2\left(1+M_0r^2\right)}\right]e^{-2H_0t}
-\rho \,, \label{5b}\\
q &=& -\left[{4M_0H_0r\over a_0\left(1+M_0r^2\right)^2}\right]
e^{-H_0t}\,. \label{5c}
\eeq
Note that the magnitude of the heat flux is a maximum at 
$r=1/\sqrt{3M_0}$, falling
to zero at the centre and as $r\rightarrow\infty$.
The causal transport equation (\ref{8}) becomes
\be
4M_0H_0r\left[\left(1+M_0r^2\right)-H_0\tau \right]=\kappa
\left(1+M_0r^2\right)^2\left[\left(1+M_0r^2\right)T\right]'\,.
\label{9}\ee
We have yet to specify $\kappa$, $\tau$ and equations of state
involving the temperature. However, we will avoid the difficult
issue of trying to introduce a microscopically motivated model, since
our primary aim is only
to show the possibility of consistent solutions.
In this spirit, we will satisfy the causal transport condition
(\ref{9}) by taking
\beq
\tau &=& \left(1+M_0r^2\right)H_0^{-1} \,,\label{10a}\\
T &=& {U(t)\over 1+M_0r^2} \,,
\label{10b}
\eeq
where $U$ is an arbitrary positive function. Equation (\ref{10a})
implies that the relaxation time is the same as the local expansion
time, which is not unreasonable.
The radially inward heat flux counters the cooling brought about by
expansion. However, one might expect that the inflationary cooling
will not be balanced or overcome by dissipative heating, and then one
would choose $\dot{U}<0$.

For any choice of $U$,
it follows from (\ref{10b}) that $T$ decreases radially outward, in
apparent contradiction to the radially inward direction of the
heat flux. However, in relativity, the inertia of heat energy gives
rise to an accelerative contribution to the heat flux. In our model,
the latter dominates the temperature gradient.
Note that, whatever the (positive)
choice of $U$ and $\kappa$, the second law of
thermodynamics will be satisfied, since it is built into the causal
theory \cite{hl}.

The model given by equations (\ref{5a})--(\ref{5c}), (\ref{10a})
and (\ref{10b}) describes
inhomogeneous inflation driven by causal heat flux. The inflationary
expansion rapidly reduces the pressure, as shown by equation
(\ref{5b}), but the heat flux counters the tendency for the energy
density to decrease, and equation (\ref{5a}) shows that the energy
density is a comoving constant ($u^\alpha\nabla_\alpha \rho=0$).
At late times, the heat flux tends to zero and the fluid approaches
a de Sitter-like equilibrium with $p=-\rho$. The deviation of the
fluid from equilibrium is measured by the covariant
dimensionless ratio
\be
{|q|\over\rho}=
\left({4M_0\over 3a_0H_0}\right)r
e^{-H_0t}
\label{11}\ee
At late times, this ratio rapidly becomes small. However for any 
fixed
finite time, the ratio grows with radius, reflecting the fact that 
$|q|$
decays less rapidly than $\rho$. Thus the fluid is only close to
equilibrium near the centre and at late times.
By contrast, in bulk viscous inflation
the fluid is far from equilibrium at all times and positions.


We have shown that in principle it is possible for causal
heat flux to drive inflation. This approach could be the basis for 
a model of inflationary expansion of bubbles in the early
universe, driven by causal energy transport processes.
Our model is of course highly simplified, and lacks some of the
physical properties that would be expected 
of a more realistic model. For example, there is
no mechanism in our simple model for achieving an exit from inflation,
and the model inflates for all time. Furthermore, the heat flux does
not homogenize the universe, as one might intuitively expect, so that
the metric does not tend to the FLRW metric asymptotically. This is
a direct consequence of the simple choice of $M(t)$ as a constant.
However, in principle it should be possible to overcome these 
drawbacks with a more sophisticated and complicated model.

\[ \]


\end{document}